\documentclass[twocolumn,prb,showkeys,showpacs,preprintnumbers,amsmath,amssymb]{revtex4}
\usepackage{graphicx}
\usepackage[large]{subfigure}
\usepackage{dcolumn}
\usepackage{bm}

\begin{document} 
\title{Magnetic interactions in transition-metal doped ZnO: An ab-initio study}
\author{Priya Gopal}
\author{Nicola A. Spaldin}%
\affiliation{%
Materials Department\\
University of California, Santa Barbara, California 93106-5050, USA\\
}%

\date{\today}

\hyphenation{%
P
rea-sons
}

\begin{abstract}
We calculate the nature of magnetic interactions in transition-metal doped ZnO using the 
local spin density approximation and LSDA+\textit{U} method of density functional theory. 
We investigate the following four cases: (i) single transition metal ion types (Cr, Mn, 
Fe, Co, Ni and Cu) substituted at Zn sites, (ii) substitutional magnetic transition metal 
ions combined with additional Cu and Li dopants, (iii) substitutional magnetic transition 
metal ions combined with oxygen vacancies and (iv) pairs of magnetic ion types (Co and Fe,
Co and Mn, etc.). Extensive convergence tests indicate that the 
calculated magnetic ground state is unusually sensitive to the k-point mesh and energy cut-off, 
the details of the geometry optimizations and the choice of the exchange-correlation functional. 
We find that ferromagnetic coupling is sometimes favorable for single type substitutional 
transition metal ions 
within the local spin density approximation. However, the nature of magnetic interactions 
changes when correlations on the transition-metal ion are treated within the more realistic 
LSDA + \textit{U} method, often disfavoring the ferromagnetic state. The magnetic configuration
is sensitive to the detailed arrangement of the ions and the amount of lattice relaxation, 
except in the case of oxygen vacancies when an antiferromagnetic state is always favored. 
\end{abstract}

\maketitle

\section{Introduction}
Dilute magnetic semiconductors (DMSs), obtained by partial replacement of the cations in 
conventional semiconductors by magnetic transition-metal ions, are of current interest as 
potential semiconductor-compatible magnetic components for spintronic applications~\cite{Ohno98}. Early studies of DMSs focussed on II-VI semiconductor hosts such as CdSe and ZnTe~\cite{Furdyna88}; 
since many transition metals adopt divalent ionic states, they therefore substitute readily for 
divalent cations such as Zn$^{2+}$ or Cd$^{2+}$. However, most II-VI DMSs are antiferromagnetic 
or have very low ferromagnetic Curie temperatures~\cite{Saito02}, rendering them unattractive for 
applications. More recently, robust ferromagnetism was observed in DMSs based on III-V semiconductors 
such as Mn-doped GaAs, in which Curie temperatures of $\sim$150 K have been achieved~\cite{Ku_Samarth03}. 
Here, the ferromagnetic coupling between the localized Mn magnetic moments is believed to be mediated 
by itinerant carriers (holes) which are introduced when divalent Mn ions replace trivalent 
galliums~\cite{Dietl00,Sanvito01} (so-called {\it carrier-mediated ferromagnetism}). Calculations
for other materials, based on such a carrier-mediated mechanism with a high concentration ($\sim$5 \%)
of holes, predicted above-room-temperature ferromagnetism for Mn-doped ZnO and GaN~\cite{Dietl00}.

In addition to the technological appeal of room-temperature ferromagnetism in ZnO-based DMSs, ZnO 
offers other desirable features as a semiconductor host. It has a direct wide band gap of 3.3 eV~\cite{LB},and therefore finds widespread use in the optoelectronics industry~\cite{Norton04,Look01,Look03}. Its
strong piezoelectricity~\cite{Hill_Waghmare00} is exploited in a variety of transducer applications~\cite{Wood_austin}
and have possible application in polarization field effect transistors~\cite{Gopal_Spaldin06}. 
And long spin coherence times, with potential spintronic applications, have recently been reported at room 
temperature in n-type ZnO~\cite{Ghosh05}.
Thus, if it could be achieved, ferromagnetic ZnO would be a highly multifunctional material with coexisting 
(and possibly coupled) magnetic, piezoelectric, optical and semiconducting properties.  

The predictions of high-temperature ferromagnetism \cite{Dietl00} spawned a large number of 
experimental~\cite{Ueda01,Coey04,Waghmare05,Fukumura:Mn1,Jung:Mn,Sharma04,Song06,Risbud03,Kittilsved,Janisch05,Thota06_nano} and computational~\cite{Min-Park-2003,Sato00,Sandratski06,Spaldin04,Patterson05,Petit06,Lee04} studies of transition-metal (TM)-doped ZnO. 
The reported experimental 
values of Curie temperature and magnetization show a large distribution, suggesting that the system is 
sensitive to preparation methods, measurement techniques, substrate choice, etc. For example, in the 
case of Mn-doped ZnO, while some experiments report above room temperature ferromagnetism~\cite{Sharma04,Heo03}, 
others report a low ferromagnetic ordering temperature~\cite{Jung:Mn,Yoon03:Mn,Kim-Choo03} or a spin-glass 
or paramagnetic behavior~\cite{Fukumura:Mn1,Dabrowski-Co-Mn,Risbud03}. Likewise, in Co-doped ZnO, there are
reports of giant magnetic moments ($\sim$ 6.1 $\mu_B$/Co)~\cite{Song06}, high ferromagnetic ordering temperatures 
with moments of $\sim$ 1-3  $\mu_B$~\cite{Coey04,Coey05,Morkoc05,Jeong05} and in some cases no ferromagnetic 
behavior at all~\cite{Ando01,Dabrowski-Co-Mn,Risbud03,Thota06_nano}.
Interestingly, the reported computational results show a similar 
spread of values; this arises in part from different 
physical approximations (choice of exchange-correlation functional, inclusion or omission of structural 
relaxations), but also from the unusual sensitivity of the magnetic interactions to the convergence
quality of the computations. (For a detailed review see Ref. \onlinecite{Janisch05}). As a result, 
in spite of the flurry of experimental and theoretical work, no 
definite conclusions have been reached regarding the nature and origin of the magnetic interactions in this 
system. 

In this work, we report our results of a systematic computational study of transition-metal-doped ZnO,
with the goal of understanding the nature and origin of the magnetic interactions. Our emphasis is on 
extracting trends along the $3d$ transition metal series, and so we first explore the effects of 
substitutional doping with a range of transition metals (Cr, Mn, Fe, Co, Ni and Cu). Next, we calculate 
the effects of additional dopant or vacancy impurities, and finally we calculate the preferred magnetic 
configurations of \emph{pairs} of different magnetic ion types. Our main finding is that the ferromagnetic 
state is generally disfavored in the most realistic calculations, and it is not strongly stabilized by
common defects likely to be found in as-grown ZnO, or easily incorporated as dopants. 

The remainder of this paper is organized as follows. In the next section, we describe our computational
and system details and outline the unusually demanding convergence behavior that we find for TM-doped
ZnO. In Sec.~\ref{single_type_TMs}, we present our calculated trends across the $3d$ series for single-type 
transition metal doping within the local spin density approximation (LSDA) and the LSDA +U method. In Sec.~\ref{defects} we investigate the influence of defects across the $3d$ series by (i) co-doping with 
Cu or Li and (ii) incorporation of oxygen additional oxygen vacancies. In Sec.~\ref{ferrimagnetic}, we 
present our results for double doping (simultaneous doping of two different TM ions) in (Zn,(Co,Fe))O and 
(Zn,(Mn,Co))O. Finally, we summarize our results in Sec.~\ref{discussion}

\section{Technicalities}

\subsection{System details}
\begin{figure}[h]
\begin{center}
\includegraphics*[width=0.15\textwidth]{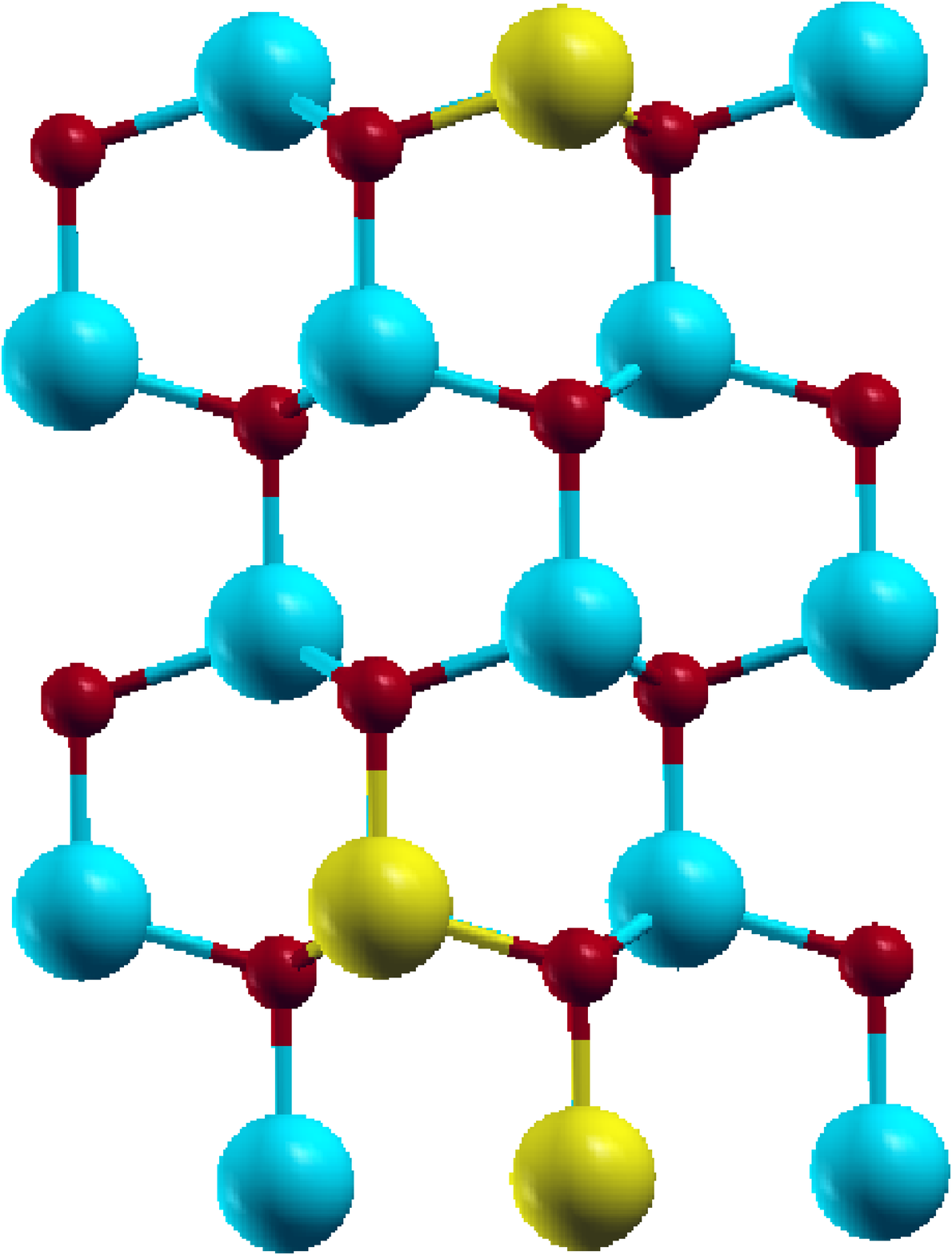}
\hspace{0.5cm}
\includegraphics*[width=0.04\textwidth]{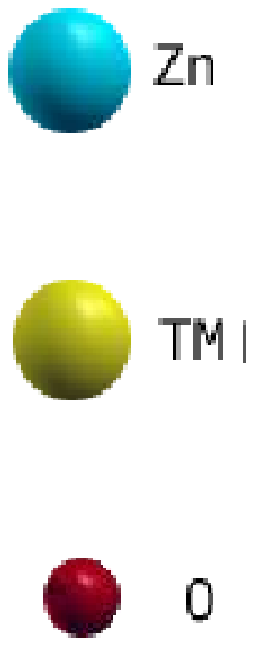}
\hspace{0.5cm}
\includegraphics*[width=0.15\textwidth]{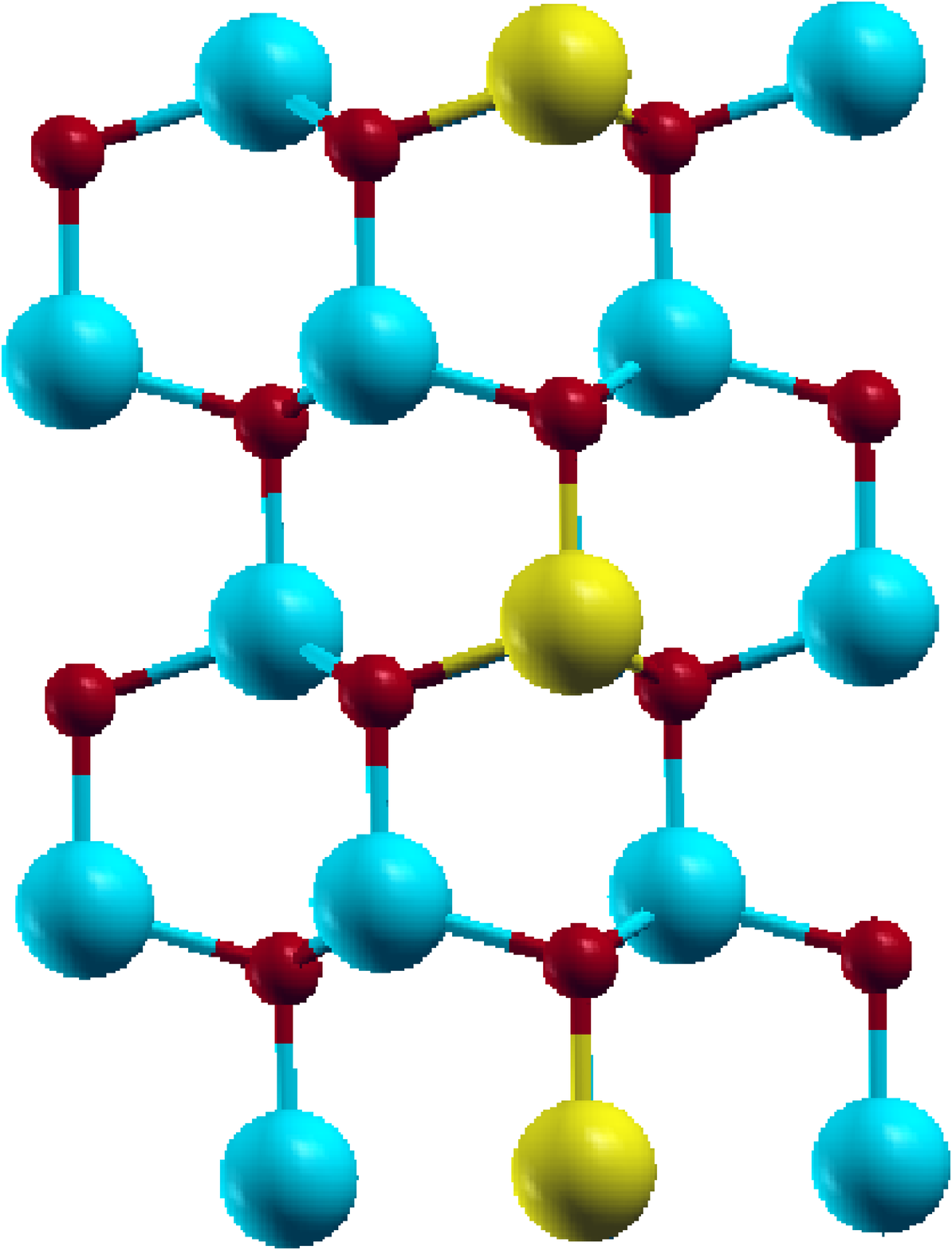}\\
(a)\hspace{4cm}(b)
\end{center}
\caption{\label{near}(Color online) Wurtzite supercell of ZnO with two TM ions in the (a) near and (b) far configuration. 
The (yellow) light-shaded spheres are the TM ions, the large (cyan) dark-shaded spheres the Zn ions, and the small (red)
black spheres the O ions.}
\end{figure}

In order to achieve realistic experimental dopant concentrations ($\sim$ 10 - 30\%), we used 
a periodic 2$\times$2$\times$2 wurtzite supercell of ZnO which consists of 32 atoms in a unit cell. 
Substitution of two Zn atoms by transition metal ions then gives a dopant concentration of 12.5\%
and allows for calculation of the relative energies of ferromagnetic (FM) and antiferromagnetic (AFM) 
orderings. We explored two spatial arrangements, \textit{near} (transition metal atoms separated by one
oxygen atom) and  \textit{far} (TM atoms separated by -O-Zn-O- ), as shown Figure~\ref{near}.
In each case, we used the energy difference between FM and AFM ordering, $\triangle$E = E$_{AFM}$-E$_{FM}$, as
an indicator of the magnetic stability (a positive $\triangle E$ implies that FM is favorable). Equating
$\triangle E$ with the thermal energy $k_B T$ suggests room temperature ferromagnetism should be achieved
for $\triangle E$ values larger than $\sim$ 30 meV. 
\subsection{Method details}
\begin{figure}
\begin{center}
\includegraphics[width=0.4\textwidth]{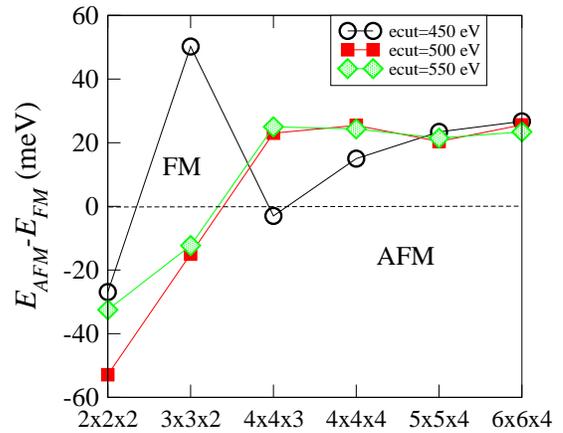}
\caption{\label{kpt-test}Plot of E$_{AFM}$-E$_{FM}$ for Zn$_{0.875}$Co$_{0.125}$O as a function of k-point grid,
and for three different plane-wave energy cut-offs (ecut).}
\end{center}
\end{figure}

Our total energy and electronic structure calculations were performed using the projector augmented wave 
(PAW)~\cite{PAW} formalism of density functional theory as implemented in the VASP package~\cite{vasp,vasp2,vasp3}. 
We used the default VASP PAW potentials with 12 valence electrons for Zn (\textit{3d$^{10}$4$s^2$}), 6 for O 
(\textit{2p$^{4}$2$s^2$}) and 2+$n$ (\textit{3d$^{n}$4$s^2$}; $n$ = 3 to 9) for the transition metals. 
We used a well converged energy cut-off of 550 eV for plane-wave expansion of the PAW's, a 4 $\times$ 4 
$\times$ 3 Gamma centered k-point grid, and the tetrahedron method with Bl\"{o}chl~\cite{PAW} corrections 
for the Brillouin Zone integrations. Note that this energy cut-off and k-point sampling are unusually
high; our convergence tests (Figure~\ref{kpt-test} indicate that qualitatively incorrect magnetic 
behavior often occurs for lower values. Similarly rigorous convergence parameters have been shown to
be required for accurate calculations for Co-doped TiO$_2$~\cite{Janisch-2006}.
For all geometry optimizations, we kept the volume of the supercell fixed to the
experimental volume (\textit{a}=6.50{\AA} and $c$=10.41 {\AA}~\cite{LB}) and relaxed all the internal 
coordinates until the Hellman-Feynman forces were less than 10$^{-3}$ eV/ \AA.
 
We approximated the exchange-correlation functional with both the local spin density approximation 
(LSDA) and the fully localized limit of the LSDA+\textit{U} method~\cite{LDA+U}. Although widely used and
well established for many properties, the LSDA is well-known to yield incorrect behavior for strongly
correlated magnetic systems, often predicting half-metallic, low spin states for systems which are 
actually insulating and high spin. For example, most TM monoxides are wide band gap antiferromagnetic 
insulators~\cite{Powell70,Messick72,Sawatsky84,Hufner92,Sawatsky92}, however the LSDA finds them 
to be either FM metals (FeO and CoO) or small-gap semiconductors (MnO and NiO)~\cite{Terakura84}. 
The LSDA+\textit{U} method extends the LSDA by explicitly adding the on-site \textit{d-d} Coulomb 
interaction, $U$, and the on-site exchange interaction, $J$, to the LSDA Hamiltonian, and usually 
gives improved results for magnetic insulators. Here we use typical values of $U=4.5$ eV and $J=0.5$ 
eV on all transition metals~\cite{AZA91,Pickett98,Coccocioni05}; although a small variation in $U$ 
and $J$ across the series is expected, our choice of constant values permits a more straightforward 
comparison. We do not modify the LSDA ZnO electronic structure, which therefore shows the usual LSDA 
underestimation of the band gap, and underestimation of the Zn $d$ state energies.

\section{Magnetic interactions in doped ZnO}
\label{single_type_TMs}

\subsection{LSDA results}

\begin{center}
\begin{figure}
\includegraphics*[width=0.4\textwidth]{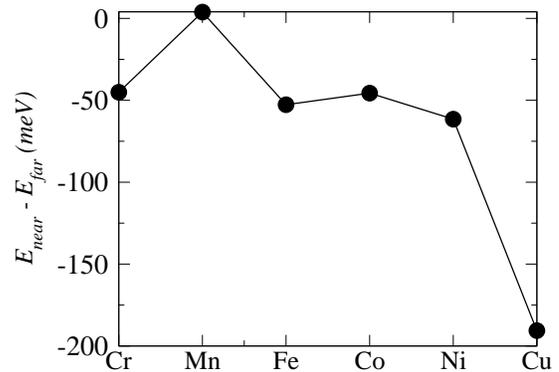}
\caption{\label{spatial-ene} LSDA energy differences ($\triangle E = E_{near}-E_{far}$) for 
substitutional TM ions. The negative energy difference indicates that the TM ions prefer to 
be in a \emph{near} spatial configuration.}
\end{figure}
\end{center}

We begin by calculating the total energies of 12.5\% TM-substituted ZnO within the local spin density
approximation (LSDA). We reiterate that we do not expect the LSDA to give accurate magnetic behavior
for this system, but present the results as a baseline for comparison with our LSDA+U results in the
next section. Keeping the volume fixed at the ZnO experimental volume, we first relax all internal 
coordinates for \textit{near} and \textit{far}  arrangements of the TM ions, and for both AFM and FM 
orderings. In figure~\ref{spatial-ene} we show our calculated energy differences between the near and 
far configurations; it is clear that in all cases (except Mn, for which the energy difference is
negligible) it is favorable for the TM ions to cluster together. 

\begin{figure*}
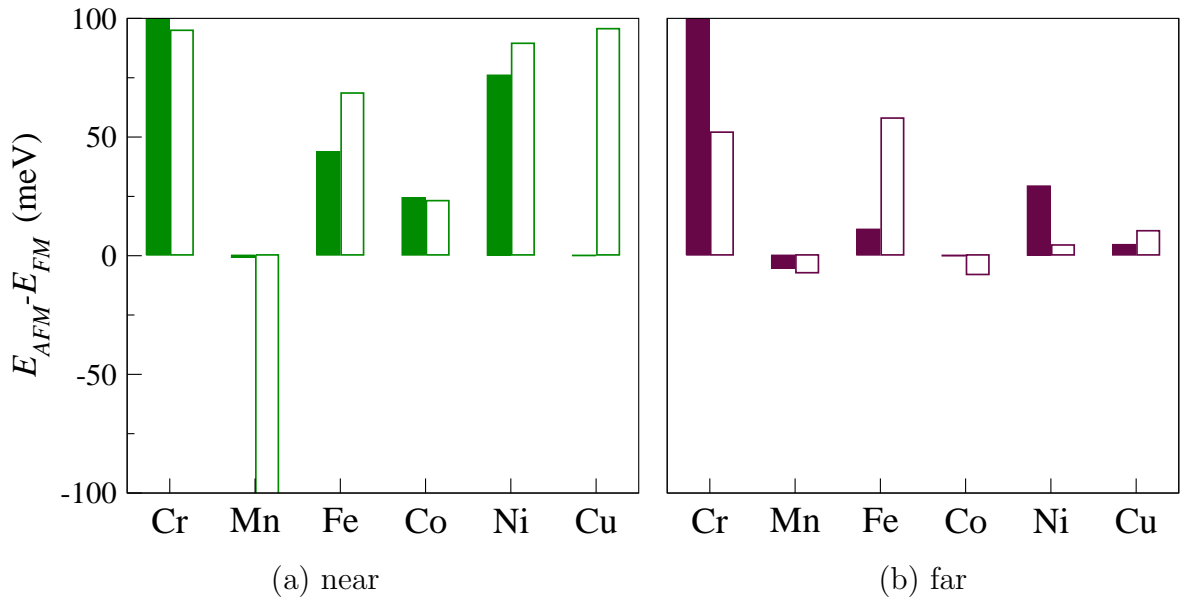

\begin{center}
      \subfigure[near]{\scalebox{0.45}{\includegraphics*{bar-lda-near.eps}}}\quad
      \subfigure[far]{\scalebox{0.45}{\includegraphics*{bar-lda-far.eps}}}\quad      
\caption{\label{LDA-fm-ene} LSDA energy differences ($\triangle E = E_{AFM}-E_{FM}$) for substitutional 
TM ions in ZnO for (a) \textit{near} and (b) \textit{far} spatial configurations. In both (a) and (b), 
the filled bars represent the case when the ionic coordinates are not relaxed and the unfilled bars 
represent the case when the ionic coordinates are relaxed. Lines at $\triangle E=0$ indicate that FM 
and AFM orderings are equivalent in energy.}
\end{center}
\end{figure*}

Figures~\ref{LDA-fm-ene}(a) and (b) show our calculated magnetic energy differences (E$_{AFM}$-E$_{FM}$) 
for our range of TM dopants in the \textit{near} and \textit{far} spatial arrangements. First we note that 
our calculated magnetic orderings without ionic relaxations, which vary considerably both with the TM
type and with the spatial arrangement, are consistent with earlier calculations in the 
literature~\cite{Sato00,Park03c,Waghmare05}. We see, however, that the strength and sign of the magnetic 
interaction is highly sensitive to the ionic relaxation, which changes both the distance between the TM 
ions (by up to $\sim$ 0.25 \AA) and the TM-O-TM angle (by up to $\pm$ 5\%). We find no clear trends in
the magnitude of relaxation, nor any direct correlation between the relaxation and the change in the magnetic 
interactions, across the TM series. We attribute the sensitivity in magnetic ordering to a subtle competition 
between antiferromagnetic superexchange (favored by 180$^\circ$ bond angles), ferromagnetic superexchange 
(favored by 90$^{\circ}$ TM-O-TM bond angles) and AFM direct exchange (favored by short TM-TM distances)~\cite{GKA1,GKA2,GKA3}.
Our results indicate that ionic relaxations must always be included in calculations to obtain meaningful results, 
and that even the generalization of relaxed positions from one TM ion to another should be applied with 
caution~\cite{Waghmare05,Park03c}. For the most realistic configurations (\emph{near}, with relaxations), 
the LSDA suggests that doping with Cr, Fe, Co, Ni and Cu  should lead to a ferromagnetic ground state.

\begin{center}
\begin{figure*}
\includegraphics*[width=0.8\textwidth]{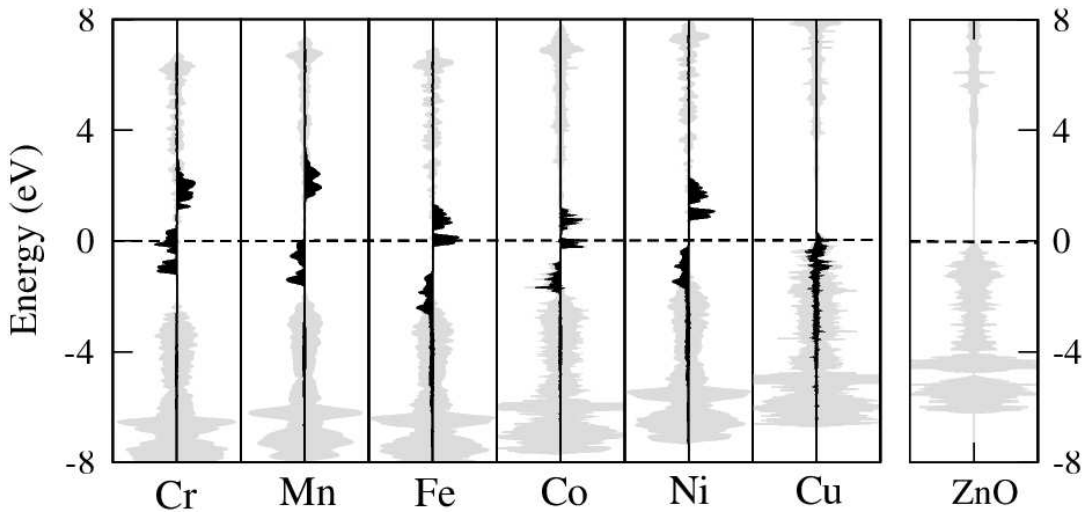}
\caption{\label{LDA-DOS} DOS and PDOS of TM-doped ZnO with the TM atoms in the \emph{near} spatial configuration calculated within the LSDA. The black shaded regions 
show the TM \textit{d} states, and the gray shaded regions show the total DOS. For clarity, the TM $d$ 
states are scaled by a factor of two. Also shown (right panel) is the DOS for undoped ZnO.}
\end{figure*}
\end{center}

Next (Figure~\ref{LDA-DOS}) we compare our calculated total densities of states (DOS) and TM $3d$  projected 
local densities of states (PDOS) for our range of TM ions in ZnO. In all cases the DOS for the FM arrangement
is shown. The total DOS is represented by the 
gray shaded region while the black shaded regions represent the \textit{3d} states of the TM impurities.
The majority ($\uparrow$) spin states are plotted along the negative $x$ direction, and the minority 
($\downarrow$) states are plotted along the positive $x$ direction. The energies are reported relative to 
the Fermi energy (E$_f$=0). For comparison, the DOS of undoped ZnO is also shown, with the Fermi energy
set to the top of the valence band.

Comparison with the ZnO DOS identifies the broad ($\sim$ 4-5 eV) band just below the Fermi energy as derived
largely from the O $2p$ states, with the narrow Zn $3d$ band just below and slightly overlapping with the 
O bands. The bottom of the conduction band is composed largely of Zn $4s$ states, and the band gap ($\sim$0.78 
eV) shows the usual LSDA underestimation. 

In all cases, the exchange-split TM \textit{3d} states form fairly narrow bands in the region of the gap. 
The exchange splitting ($\sim$2 eV) is consistently larger than the crystal field splitting ($\sim$0.5 eV)
which splits each spin manifold into lower energy doubly degenerate $e$ and higher energy triply degenerate 
$t$ states. We observe the following trends, which are consistent with the crystal chemistry of $3d$ 
transition metal oxides:
\begin{itemize}
\item The energy of the TM $3d$ states relative to the top of the valence band shifts down in energy 
on moving right across the $3d$ series (from Cr to Cu). As a consequence, the Cr states are somewhat 
hybridized with the bottom of the conduction band (this is likely a result of the LSDA underestimation 
of the ZnO band gap), the Fe and Co states are largely mid-gap, and the Cu states 
are hybridized with the top of the valence band.
\item The calculated DOSs (Figure~\ref{LDA-DOS}) are consistent with the Hund's rule predictions for 
isolated TM ions, with only majority-spin ($\uparrow$) states occupied for Cr and Mn, and 
occupation of the minority-spin ($\downarrow$) states increasing Fe to Co to Ni to Cu.
Also the exchange splitting decreases on moving to the right across the $3d$ series. 
\item The $(d^5)^{\uparrow}$ Mn$^{2+}$ ion, and $(d^5)^{\uparrow}$  $(d^2)^{\downarrow}$ Co$^{2+}$ ions 
lead to insulating behavior; the other ions have partially filled $d$ manifolds and a finite density of 
states at the Fermi level.
\item The variation in magnetic moments across the $3d$ series reflects the decrease in exchange
splitting, and the increase in number of electrons, with the local magnetic moments (obtained by integrating up to a Wigner-Switz radius of 1.4 \AA)~\cite{Shannon76} on the TM ion 
as follows (in units of $\mu_B$):  3.04 (Cr), 4.31 (Mn), 3.51 (Fe), 2.44 (Co), 1.49 (Ni) and 0.55 (Cu). 
\end{itemize}

These LSDA results are consistent with earlier LSDA calculations~\cite{Sato00,Spaldin04,Waghmare05}, and 
also with some experimental reports~\cite{Sharma04}. Since the LSDA is known to underestimate the band gaps 
and exchange splittings in magnetic systems, we withold a detailed analysis of the band structures until 
the next section, where we repeat our calculations with the inclusion of correlations at the level of the 
LSDA + \textit{U} method. 

\subsection{LSDA +\textit{U}}

Next we repeat the suite of calculations described above, with the correlation on the $3d$ TM ions treated 
within the LSDA +U scheme~\cite{LDA+U}. Here we present results obtained using typical values of 
\textit{U} = 4.5 eV and \textit{J} = 0.5 eV; in the appendix we discuss the \textit{U}-dependence of our 
calculated properties. 

\begin{center}
\begin{figure*}
\includegraphics*[width=0.8\textwidth]{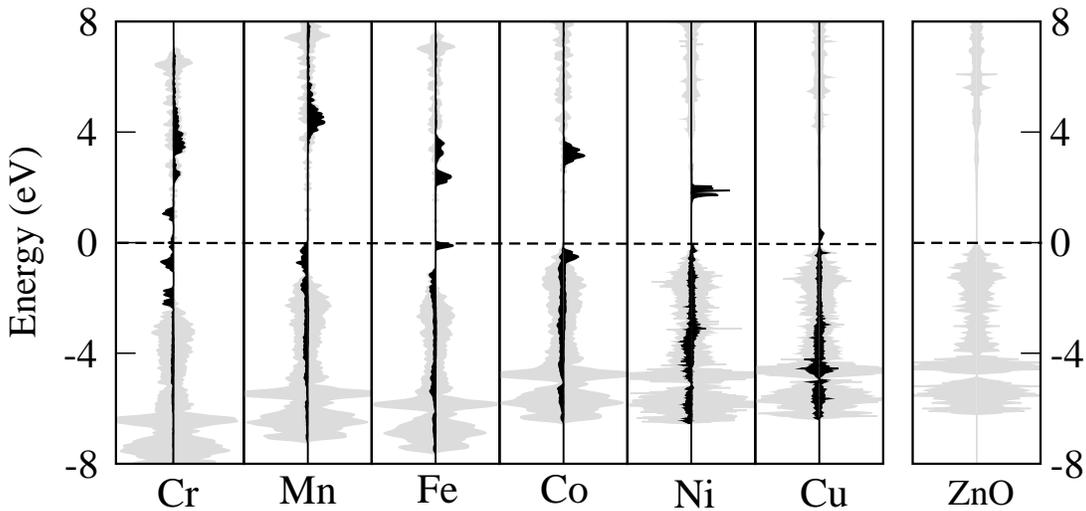}
\caption{\label{LDAU-dos}DOS and PDOS of TM-doped ZnO with TM atoms in the \emph{near} spatial configuration calculated within the LSDA +\textit{U} method. 
The black shaded regions show the TM \textit{d} states, and the gray shaded regions show the total DOS. 
For clarity, the TM $d$ states are scaled by a factor of two. Also shown (right panel) is the DOS for 
undoped ZnO.}
\end{figure*}
\end{center}
\begin{figure*}
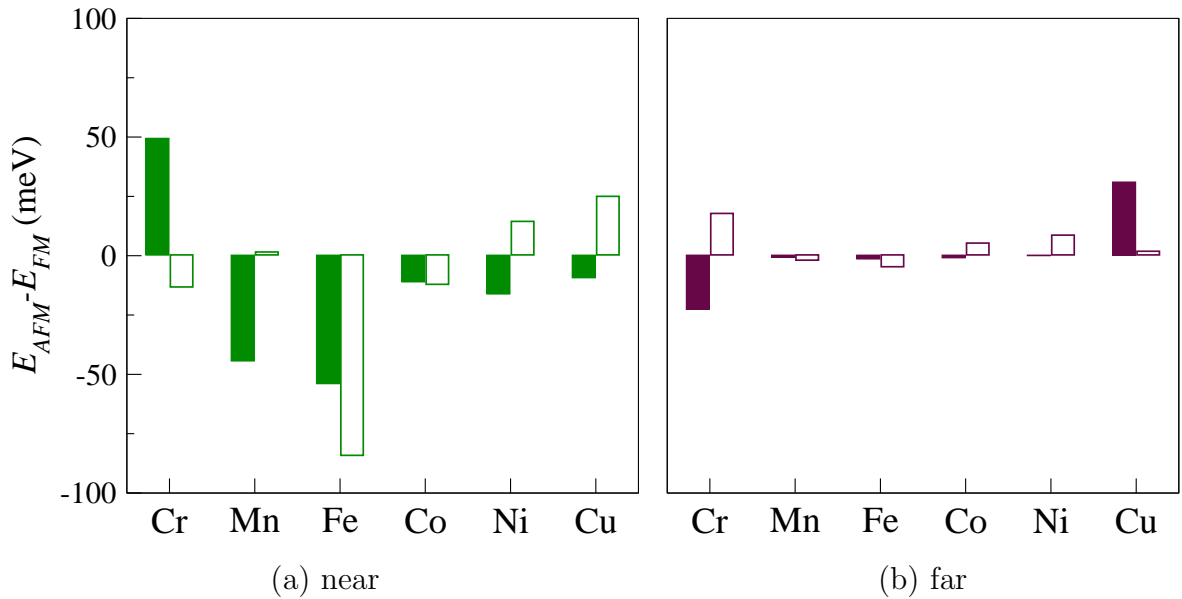

\begin{center}
\mbox{
        \subfigure[near]{\scalebox{0.45}{\includegraphics*{bar-ldau-near.eps}}}\quad 
      \subfigure[far]{\scalebox{0.45}{\includegraphics*{bar-ldau-far.eps}}}\quad     } 
\caption{\label{LDAU-tm-ene}LSDA \textit{+U} energy difference, $\triangle$E=E$_{AFM}-E_{FM}$, for substitutional TM ions 
in ZnO represented for (a) \textit{near} and (b) \textit{far} spatial configurations. The filled bars show the energies
when the ions are not relaxed and the shaded bars give the values for when relaxations are included. Lines at $\triangle$E=0 
indicate that FM and AFM orderings are equivalent in energy.}
\end{center}
\end{figure*}

First, in Figure~\ref{LDAU-dos} we present our calculated LSDA+U DOSs of TM-doped ZnO. Although all the trends across
the series remain the same as in the LSDA, the addition of the Hubbard \textit{U} term, as expected, causes an increased 
splitting between the filled and empty orbitals in the TM $d$ manifold. The exchange splitting is now increased to 
$\sim$4 eV, and the crystal-field splitting increases to $\sim$1 eV. The TM $d$ states, which were localized in the gap 
in the LSDA, now shift down in energy and hybridize strongly with the O $2p$ states. Comparing the DOS of Fe and Ni, we observe a transition from half-metallic to insulating when going from the LSDA to LSDA +U. This is directly reflected in the magnetic interation strength. For e.g., in Fe and Ni, the long-range interaction vanishes and the interaction strength decreases. 
We also calculated the magnetic moments on the TM ion as follows (in units of $\mu_B$): 3.40 (Cr), 4.41 (Mn), 3.54 (Fe), 2.56 (Co), 1.63 (Ni) and 0.65 (Cu). The magnetic moments are slightly larger compared to the LSDA values.  
Next we compare the total energies of the FM and AFM magnetic orderings in the \textit{near} and \textit{far} spatial 
configurations. In all cases, all the ionic coordinates are relaxed within the LSDA +U scheme.
Figures~\ref{LDAU-tm-ene}(a) and (b) show the energy difference between the FM and AFM states,  ($E_{AFM}-E_{FM}$), for 
\textit{near} and  \textit{far} arrangements for the various TM ions.
As in the LSDA case, $\Delta E$ shows a strong dependence on the spatial arrangement of the dopant ions, as well as on the 
extent of ionic relaxations included in the calculation. In all cases, the sign and magnitude of the interactions are 
strikingly different from those obtained within the LSDA, with a general trend to reduced ferromagnetism. For the energetically 
favorable \textit{near} case, only Ni and Cu dopants show a tendency to ferromagnetic ordering when ionic relaxations are
included in the calculation.

\section{Influence of defects}
\label{defects}

\subsection{Possible \textit{p}-type dopants; Cu and Li with TM in ZnO}

\begin{center}
\begin{figure}
\includegraphics*[width=0.45\textwidth]{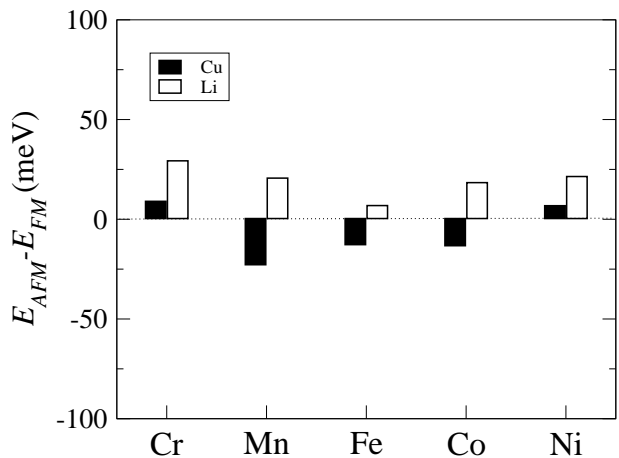}
\caption{\label{ZnO-Li-Cu-ene} E$_{AFM}-E_{FM}$ for TM-doped ZnO with 6.25\% Cu/Li.}
\end{figure}
\end{center}

Given that our calculations containing single types of subsitutional dopants are unable to reproduce the experimentally 
reported ferromagnetism, we next search for other impurities that could mediate the ferromagnetic ordering. Since the 
original predictions of high Curie temperature in Mn-doped ZnO~\cite{Dietl00} assumed high hole concentrations, we first 
include additional dopants which are likely to introduce $p$-type carriers. Some experimental~\cite{Han02} and theoretical 
studies~\cite{Park03c} have suggested that Cu could provide holes when doped into ZnO.  Cu has an electronic configuration of  
3$d^9$4$s^2$, and has a tendency to form Cu$^{1+}$ ions with a 3$d^{10}$4$s^0$ electron configuration, and an ionic radius 
(0.60 \AA) close to that of Zn$^{2+}$. Thus it is reasonable to assume that substitution of Zn with Cu could yield a 
Cu$^{1+}$ configuration with an associated hole. Indeed, earlier LSDA calculations on Cu-doped ZnO found a band
structure consistent with this model, and an enhanced tendency to ferromagnetism \cite{Spaldin04}. Lithium has also
been suggested as a possible \textit{p}-type dopant in ZnO~\cite{Ceder00}. Li has an electronic configuration of 
1s$^2$, 2s$^1$, and, like Cu, tends to form a singly charged
cation ($1s^2 2s^0$) with an ionic radius (0.59 \AA) close to that of Zn$^{2+}$. 
In this section, we calculate the effects of co-doping TM-doped ZnO with Cu or Li, again within the LSDA+U method, using
the same convergence and $U$ and $J$ parameters as described above. Our calculated $\Delta E$ values are shown in 
figure~\ref{ZnO-Li-Cu-ene}. In all cases the concentration of the Cu or Li was 6.25\% (one additional dopant in 16 cations),
and the additional dopant was placed far away from the TM ion in ZnO.  It is clear that, in contrast with earlier LSDA 
calculations, co-doping with Cu within the LSDA+U method does not strongly enhance the tendency towards ferromagnetism,
yielding an antiferromagnetic ground state in most cases, and a weakly ferromagnetic state for Cr and Ni. 
Co-doping with Li, however, is more favorable, and in some cases should lead to above-room-temperature ferromagnetism.
This is particularly intriguing in light of recent reports of ferroelectricity in Li-doped ZnO~\cite{Li-ferro:98, Li-ferro:01}, suggesting the 
possibility of engineering a multiferroic material with simultaneous ferromagnetism and ferroelectricity.
However, none of our calculated Li-doped materials is insulating, a requirement for ferroelectricity.
 
\subsection{O vacancies: Zn(TM,V$_o$)O}
Next, we combine substitutional TM ions with oxygen vacancies, which are believed to be the most common native defects 
in ZnO~\cite{Zhang01,Ceder00,Look05}. There is some experimental evidence associating the presence of oxygen vacancies with the
existence of ferromagnetism in TM-doped ZnO. For example, Venkatesan~\cite{Coey04} \textit{et al.} reported a correlation
between the magnitude of Co magnetic moments and the oxygen partial pressure during annealing, with higher oxygen partial
pressure reducing the amount of magnetization. In addition, Kittilsved 
\textit{et al}~\cite{Kittilsved} report the observation of room-temperature ferromagnetism in Co-doped ZnO nanocrystals
with oxygen-containing surfaces, and propose that oxygen vacancies mediate the magnetic interactions. A couple of experiments also observe a change in the magnetization in samples annealed in reduced atmospheric pressures and they attribute it to oxygen vacancies~\cite{Naeem06}.
As a result of this reported correspondence between the presence of oxygen vacancies and the existence of ferromagnetism,
a model has been proposed in which ferromagnetism is mediated by carriers in a spin-split impurity band derived from
extended donor orbitals \cite{Coey04,Coey05}. The validity of the model rests on the formation of an 
oxygen-vacancy-derived donor impurity band close to the ZnO conduction band edge, which hybridizes with the 
spin-polarized TM $3d$ band. If the Fermi energy lies within this hybrid oxygen vacancy/TM $3d$ spin-polarized
band, a carrier-mediated ferromagnetism should be favorable.
Here we calculate the relative energies of the oxygen impurity levels and the transition metal $3d$ states
across the TM series, in order to investigate the applicability of this model. 
Indeed, by simple charge neutrality arugments, removal of a neutral oxygen atom should leave two unbonded electrons associated
with the vacancy, provided that the Zn remains in a Zn$^{2+}$state.  We note, however, that recent first-principles 
calculations~\cite{Janotti05} found oxygen vacancies to be deep donors in ZnO, and therefore unavailable for 
carrier mediation. 

\begin{center}
\begin{figure}
\includegraphics*[width=0.45\textwidth]{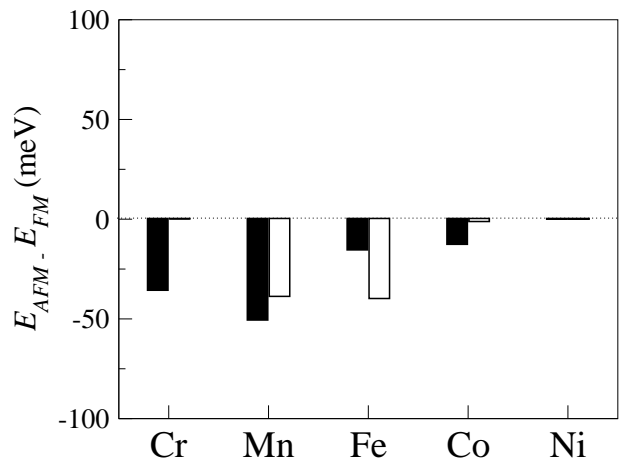}
\caption{\label{defect-ovac} $E_{AFM}-E_{FM}$ for the range of TM ions in the presence of an oxygen vacancy. The shaded and 
unshaded bars represent the cases when the TM ions are \textit{near} and \textit{far} from each other respectively. In both 
cases the oxygen vacancy is placed as far as possible from the TM ions.}
\end{figure}
\end{center}

We introduce a single O vacancy as far as possible from the TM ions which are again in both the \textit{near} and \textit{far} 
configurations. The energy difference ($E_{AFM}-E_{FM}$) is plotted in Figure~\ref{defect-ovac}. 
We see that the AFM state is stable in all cases, consistent with previous 
computations for Co-doped ZnO with a neutral oxygen vacancy~\cite{Spaldin04,Patterson05,Waghmare05}. 

\begin{figure*}[htbp]
\begin{center}
\includegraphics*[width=0.8\textwidth]{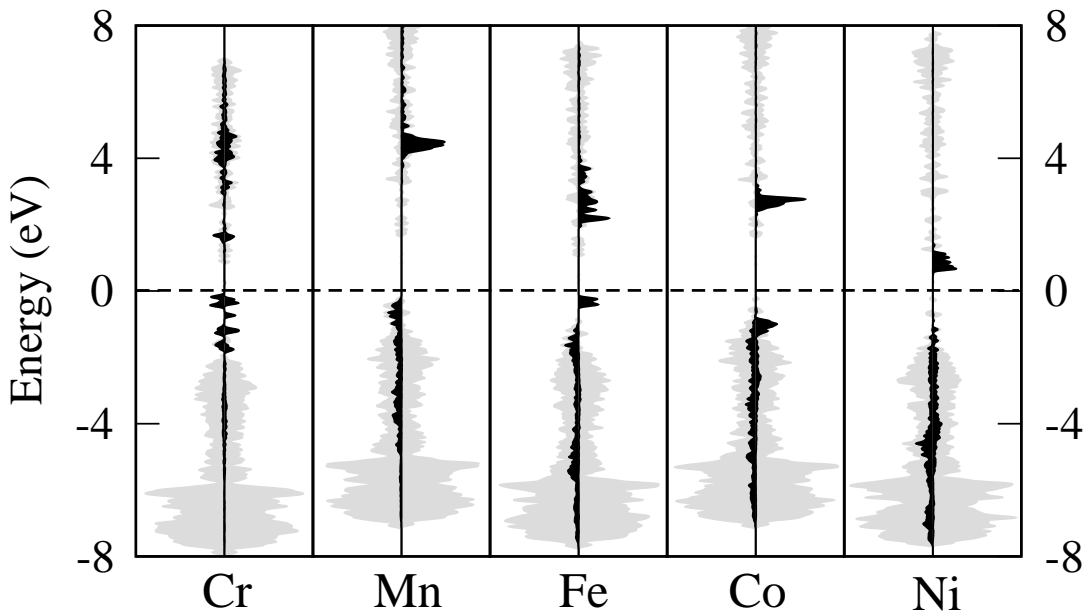} 
\caption{\label{ovac-tm-dos}DOS and PDOS for TM-doped ZnO (\emph{near} spatial arrangement and FM
state) in the presence of oxygen vacancies.}  
\end{center}
\end{figure*}
In Figure~\ref{ovac-tm-dos} we show our calculated DOSs for the entire series. Note that the defect concentration
used in our calculations is higher than likely attainable experimentally ($\sim$1-2\%), but that the relative band 
alignments will be largely unchanged by the increased concentration. We see a shift of the Fermi energies relative
to those in the absence of oxygen vacancies, but no other striking changes in  band structure. In particular,
it is clear that the Fermi level does not lie in a hybridized TM $3d$-O $2p$ impurity band for any case. 
Therefore our band structures are not consistent with those required to mediate ferromagnetism within the
model proposed above; they are however consistent with our calculated AFM ground states. 

The calculated magnetic moments (in units of $\mu_B$) for each TM ion in the presence of the oxygen vacancy are 
3.54 (Cr), 4.36 (Mn), 3.66 (Fe), 2.64 (Co) and 1.68 (Ni), largely unchanged from the LSDA+U values without the
oxygen vacancy. This indicates that the two electrons from the 
oxygen vacancy are localised and do not influence the occupancy of the TM $d$ states. 
Finally, we point out that we have considered only neutral oxygen vacancies here. A recent first principles study 
suggests that positively charged oxygen vacancies might be more successful for mediating ferromagnetism in Co-doped ZnO~\cite{Patterson05}. 

\subsection{Other defects}

\begin{center}
\begin{figure}
\includegraphics*[width=0.5\textwidth]{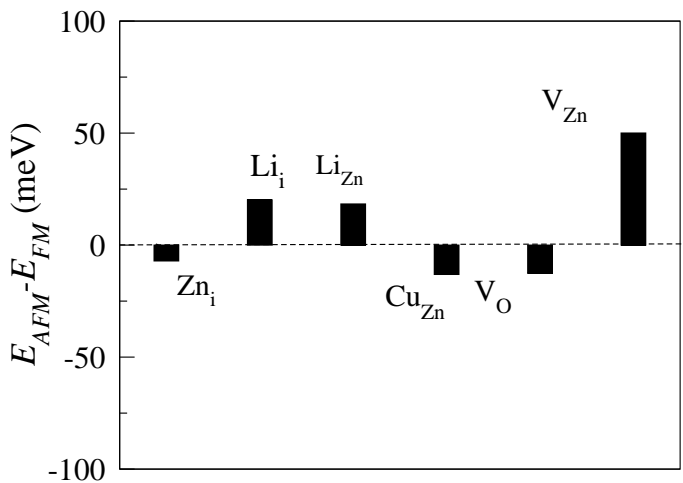}
\caption{E$_{AFM}-E_{FM}$ for ZnO doped with Co and a range of other defects.
\label{other_defects}} 
\end{figure}
\end{center}

Next, we study the effect of a range of experimentally plausible defects -- Zn vacancies (V$_{Zn}$), octahedral 
Zn interstitials (Zn$_i$), TM interstitials (Co$_i$) and Li interstitials (Li$_i$) -- on the magnetic interactions. 
Rather than scanning the entire range of transition metal dopants, we use Co-doped ZnO as our test system, since 
it is the most widely studied both experimentally~\cite{Ueda01,Yang02,Cho02,Coey04,Janisch05} and 
theoretically~\cite{Spaldin04,Patterson05,Park03c,Chang04,Waghmare05}. Figure~\ref{other_defects} shows our
calculated $\Delta E = E_{AFM} - E_{FM}$ for the \emph{near} spatial arrangement, with the vacancies 
placed as far as possible from the Co ions. Our previously reported values for substitutional Li (Li$_{Zn}$)
and oxygen vacancies (V$_O$) are also shown for completeness.

Perhaps the most important result is that, as in the case of substituional Li reported above, 
interstitial Li also stabilizes the ferromagnetic state. This is significant since Li impurities
are likely to be incorporated both in substitutional and interstitial sites during growth, and
avoiding competing antiferromagnetic interactions is desirable. We also see that Zn vacancies
are favorable for ferromagnetism, consistent with previous reports in the 
literature~\cite{Spaldin04,Waghmare05,Patterson05}. The other defects studied do not stabilize 
the ferromagnetic state. 

\section{Simultaneous doping of Co with other TM ions}
\label{ferrimagnetic}

Finally, we calculate the properties of ZnO doped with both Co and an additional TM ion (Mn or Fe). 
There has been some experimental work on such co-doped ZnO-based DMSs. Cho~\cite{Cho02} \textit{et al.} 
reported room-temperature magnetism with a saturation magnetization of 5.4 emu/g (corresponding to $\sim$ 
1$\mu_B$ for CoFe pair) for Zn$_{1-x}$(Fe$_{0.5}$,Co$_{0.5}$)$_{x}$O (x=0.15) films fabricated by reactive 
magnetron co-sputtering. Han~\cite{Han02} \textit{et al.} also reported room temperature magnetism for  
Zn$_{1-x}$(Fe$_{1-y}$,Cu$_{y}$)$_{x}$O bulk samples with T$_C$ $\sim$550K. On the theory side, Park and
Min~\cite{Min-Park-2003} 
calculated the electronic structures and magnetic properties of (Fe,Co) and (Fe,Cu) doped 
ZnO at 12.5\% doping. They reported a tendency to
form FM Fe-O-Cu clusters, and argued that ferromagnetism should arise from double exchange, whereas the
absence of clustering found for (Fe,Co)-doped ZnO would require a different mechanism for ferromagnetism. 

\begin{figure}
\begin{center} 
\includegraphics*[width=0.3\textwidth]{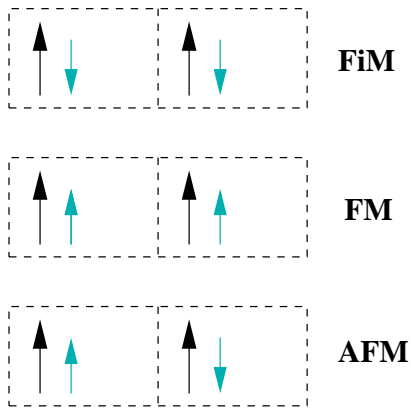} 
\caption{\label{double-TM}(Color online) A schematic of the different magnetic orderings investigated in this work. The black 
arrows represent the magnetic moments of the Co ions while the blue (gray) arrow represent the magnetic moment of the other TM ion (Mn or Fe).} 

\end{center}
\end{figure}

Here we explore the possibility of ferr{\it i}magnetic (FiM) ordering, in which the magnetic moments of one TM
ion type are antiparallel to those of the other TM ion type, but a net magnetization arises from incomplete
cancellation of the magnetic moments. For each system (Zn(Co,Fe)O, Zn(Co,Mn)O) we calculate 
the total energies of the FM, AFM and FiM magnetic orderings as shown in Figure~\ref{double-TM} with TM
ions of different type in the \emph{near} arrangement.
We use a 4$\times$4$\times$2 wurtzite supercell with 64 atoms (double the size of the previous studies), which
allows the permutations shown in Figure~\ref{double-TM} at a TM ion concentration of 12.5\%. 
Within this constrained spatial arrangement, we find that in both systems the FiM ordering is the most
stable arrangement by 200 meV for (Zn,(Co,Fe))O, and by 57 meV for (Zn,(Co,Mn))O; the corresponding magnetic 
moments per supercell are 1.11 $\mu_B$ for (Zn,(Co,Fe))O and 1.98 $\mu_B$ for (Zn,(Co,Fe))O which agrees well with the experimentally reported values~\cite{Cho02}.  
While ferrimagnetism overcomes the problem of cancellation of magnetic moments by superexchange-driven
antiparallel alignment, it of course requires that ions of different types cluster together while those
of the same type remain distant; this might be difficult to achieve experimentally. 

\begin{figure}
\begin{center} 
\includegraphics*[width=0.4\textwidth]{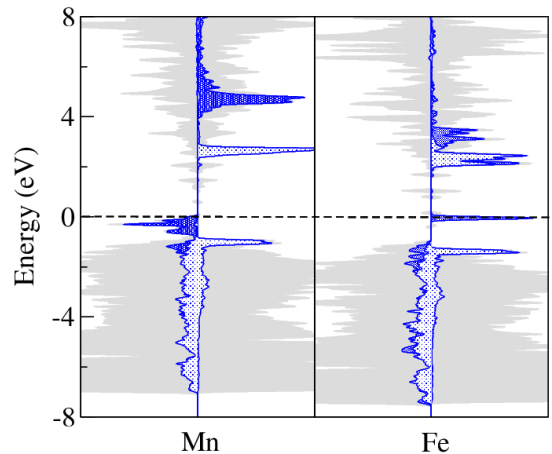} 
\caption{\label{double-fim-dos}DOSs and PDOSs of $3d$ states for ferrimagnetic Zn$_{0.875}$(Co$_{0.5}$Mn$_{0.5}$)$_{0.125}$O
and  Zn$_{0.875}$(Co$_{0.5}$Fe$_{0.5}$)$_{0.125}$O. The blue shaded region represents the $d$ states of the Co ions 
while the red shaded region represents the $d$ states of the TM ions.}
\end{center}
\end{figure}

For completeness, we show the DOSs of FiM Zn$_{0.875}$(Co$_{0.5}$M$_{0.5}$)$_{0.125}$O (M = Mn, Fe) in 
Figure~\ref{double-fim-dos}. The (Zn,(Mn,Co)) DOS is almost an exact superposition of the separate (Zn,Mn)O and (Zn,Co)O DOSs. Interestingly, in the (Zn,(Fe,Co))O system, we get a metallic state because of the narrow band spin polarized impurity band at the Fermi level unlike either (Zn,Fe)O or (Zn,Co)O. The Coey model described in the previous section could be appropriate in this picture where the donor impurity band of one the transition metal atom (in this case Fe) lies close to the conduction band edge and overlaps with the Fermi level. Thus we have a stable ferrimagnetic state. 

\section{Summary}
\label{discussion}

In summary, we have performed a systematic study of the magnetic behavior of TM-doped ZnO for a range of TM 
ions and defects. Our main result is the absence, in general, of a tendency for pairs of TM ions substituted 
for Zn to order ferromagnetically; in most cases AFM ordering is more favorable. Ferromagnetic ordering of TM 
ions is not induced by the addition of substitutional Cu impurities nor by oxygen vacancies. Incorporation of
interstitial or subsitutional Li is favorable for ferromagnetism, as are Zn vacancies. On a technical note, 
we find that the calculated magnetic behavior is strongly dependent both on the computational details (with 
ferromagnetism disfavored by improved convergence) and on the choice of exchange-correlation functional (with 
ferromagnetism disfavored by the more appropriate LSDA+U method). This observation explains the large spread
of computational results in the literature. 

\vskip 2cm
\paragraph*{Acknowledgments}
This work was funded by the Department of Energy, grant number DE-FG03-02ER4598, and was partially supported
by the MRSEC Program of the National Science Foundation under Award No. DMR05-20415. The authors thank Dr. Rebecca Janisch for useful discussions.

\appendix*
\section{Dependence of properties on choice of \textit{U}: the example of Co-doped ZnO}

\begin{figure}[h]
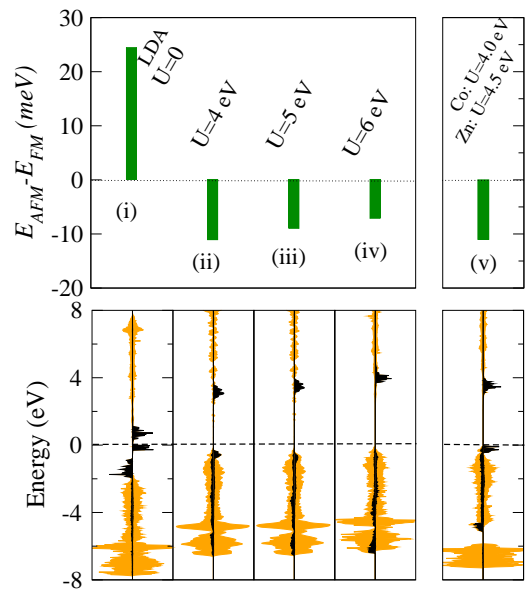

\includegraphics*[width=0.38\textwidth]{zcoo-u-bar.eps}\\
\hspace{0.15cm}\includegraphics*[width=0.372\textwidth]{fm-u.eps}\\
\caption{\label{LDAU-zcoo} Upper panel: Total energy difference (E$_{{AFM}-{FM}}$) for 12.5\% Co in ZnO for various \textit{U} values on 
the Co $d$ states. The last bar gives this energy difference with \textit{U} = 4 eV on both Zn and Co $3d$ states. The gray 
shaded and the black shaded bars show the energy differences for the \textit{near} and \textit{far} cases. 
Lower panel: Total DOS of (Zn,Co)O and PDOS of Co \textit{d} states for a range of \textit{U} values. The light shaded regions represent 
the total DOS while the black shaded regions correspond to the \textit{3d} states of Co ion.}
\end{figure}

Here we illustrate the sensitivity of our results to the magnitude of the \textit{U} parameter with calculations
of the energy difference, $E_{AFM}-E_{FM}$ and total and orbital-resolved densities of states, as a function of 
\textit{U} in the (Zn,Co)O system. The results are plotted in Figure~\ref{LDAU-zcoo}. 
Figure~\ref{LDAU-zcoo}: (upper panel) shows $E_{AFM}-E_{FM}$ for the LSDA case (with \textit{U}=0, (i)), for \textit{U} on the Co $d$
states ranging from 4 to 6 eV (points (ii), (iii) and (iv)) and with a \textit{U} of 4 eV on both the Co and Zn $d$ states (v). 
Two conclusions can be drawn from the plot: First, the sign of the 
magnetic interaction changes from LSDA to LSDA+\textit{U} in the \textit{near} case making the AFM ordering more stable. The 
AFM remains stable for a range of \textit{U} values. Second, the \textit{U} on the Co ion dominates the magnetic interaction;
the energy differences in cases  (ii) and (v) are very close, indicating that correlations on the Zn \textit{d} states do not 
significantly affect the magnetic interactions. 
Figure~\ref{LDAU-zcoo} (lower panel) shows the DOS and Co $d$ PDOS in Co-doped ZnO for a range of \textit{U} values. We see that, as
expected, the exchange splitting between occupied majority states, and unoccupied minority states increases with \textit{U};
as a consequence the majority occupied states move down into the valence band and hybridize more strongly with the O $2p$ 
states as \textit{U} is increased. Adding a \textit{U} of 4.5 eV on the Zn also lowers the energy of the Zn $d$ states.
%
\newpage
\bibliographystyle{apsrev}
\bibliography{Lit_1,Lit_ZnO,Lit_2}
\end{document}